\documentclass[a4paper,scriptaddress,twocolumn,prb,showkeys]{revtex4}

	\usepackage{amsmath}
	\usepackage{makeidx}
	\usepackage{amsfonts}
	\usepackage[ansinew]{inputenc}
	\usepackage[usenames,dvipsnames]{pstricks}
	\usepackage{subfigure}
	\usepackage{epsfig}
	\usepackage{pst-grad} 
	\usepackage{pst-plot} 

\makeindex

\begin{document}

\title{Epitaxial growth and superconducting anisotropy of PbSr$_{2}$Y$_{1-x}$Ca$_{x}$Cu$_{2}$O$_{7+\delta}$ thin films}

\author{S. Komori}
\email{komori@sk.kuee.kyoto-u.ac.jp}
\affiliation{Department of Electronic Science and Engineering, Kyoto University, Katsura, Nishikyo-ku, Kyoto 615-8510, Japan}

\author{R. Inaba}
\affiliation{Department of Electronic Science and Engineering, Kyoto University, Katsura, Nishikyo-ku, Kyoto 615-8510, Japan}

\author{K. Kaneko}
\affiliation{Photonics and Electronics Science and Engineering Center, Kyoto University, Katsura, Nishikyo-ku, Kyoto 615-8520, Japan}

\author{S. Fujita}
\affiliation{Photonics and Electronics Science and Engineering Center, Kyoto University, Katsura, Nishikyo-ku, Kyoto 615-8520, Japan}

\author{I. Kakeya}
\affiliation{Department of Electronic Science and Engineering, Kyoto University, Katsura, Nishikyo-ku, Kyoto 615-8510, Japan}

\author{M. Suzuki}
\affiliation{Department of Electronic Science and Engineering, Kyoto University, Katsura, Nishikyo-ku, Kyoto 615-8510, Japan}


\begin{abstract}
Thin films of single-crystal Pb$_{1-y}$Bi$_{y}$Sr$_{2}$Y$_{1-x}$Ca$_{x}$Cu$_{2}$O$_{7+\delta}$ (PbBi1212) were grown on SrTiO$_{3}$ (100) substrates by a two-step growth technique 
in which an amorphous film is annealed at 970~$^\circ$C in a closed ceramic container prepared using the same material as the film. 
We find that PbBi1212 exhibits superconductivity when the Ca concentration $x$ exceeds 0.3. 
The effective number of holes per Cu atom $n_{\rm eff}$ is well described as $n_{\rm eff}=0.34x$. 
The highest onset temperature for the superconducting transition attained in the present study is 65~K.
The resistivity measurement in a magnetic field reveals that the coherence lengths of Pb1212 ($y=0$) 
are approximately 25 and 2.7~\AA\, along the $ab$ plane and the $c$ axis, respectively. 
\end{abstract}


\maketitle

\section{Introduction}
The family of 1212 cuprates has single block layers and involves capable materials for superconducting wires 
because their anisotropy is less than that of the 2212 and 2223 families, which have double block layers bonded by van der Waals forces. 
The crystal structure of PbSr$_{2}$CaCu$_{2}$O$_{7+\delta}$ (Pb1212) is very similar to that of TlBa$_{2}$CaCu$_{2}$O$_{7-\delta}$ (Tl1212) and HgBa$_{2}$CaCu$_{2}$O$_{6+\delta}$ (Hg1212), 
both of which have a critical temperature $T_{\rm c}$ greater than 100~K. 
Pb1212 is expected to be less anisotropic than Tl1212 and Hg1212 because the ionic radius of the element account for the block layers is the smallest in the series ($r_{\rm Hg}$ $\approx$ $r_{\rm Tl}$ $>$ $r_{\rm Pb}$). 
Among the Pb1212 compounds, Pb$_{1-y}$Bi$_{y}$Sr$_{2}$Y$_{1-x}$Ca$_{x}$Cu$_{2}$O$_{7+\delta}$ (PbBi1212)\,\cite{Bauer,Frank,Zoller,Schneider} and Pb$_{1-y}$Cu$_{y}$Sr$_{2}$Y$_{1-x}$Ca$_{x}$Cu$_{2}$O$_{7+\delta}$ (PbCu1212)\,\cite{Maeda,Hughes} 
are of specific importance for industry because they mainly consist of the ubiquitous elements Pb and Ca rather than the more exotic Bi and Y, which are currently used for high-$T_{\rm c}$ superconducting wires. 
However, for Pb1212, the superconducting properties such as anisotropy and critical current are inadequately understood because of the difficulty of obtaining single-crystal samples of this material. 
This lack of reliable samples has even led to claims that PbBi1212 is nonsuperconducting 
and the reported superconductivity of PbBi1212 ($T_{\rm c}$ = 92~K\cite{Frank}) is attributed to the secondary phase of Bi$_2$Sr$_2$CaCu$_2$O$_{8+\delta}$ (Bi2212). 
Therefore, this study aims to verify the superconductivity of PbBi1212 by growing single-crystal PbBi1212 epitaxial films.  

Growing thin films of cuprates that incorporate Pb and Bi is extremely challenging 
because their vapor pressure is very high to maintain the chemical composition of the film at the growth temperature of approximately 700~$^\circ$C. 
Karimoto and Naito succeeded in growing an epitaxial film of PbSr$_{2}$CuO$_{5+\delta}$ by molecular beam epitaxy\,\cite{Karimoto}. 
However, they also reported that the method does not work well for the growth of Pb1212, 
because the growth temperature is limited due to the volatility of Pb\,\cite{Naito}. 
Although we attempted to grow PbBi1212 by conventional $in$ $situ$ sputtering, 
we never obtained single-phase films of PbBi1212 because of the re-evaporation of Pb and Bi from the substrate. 

In the present paper, we describe a two-step growth technique\,\cite{Yan} that allows us to maintain the chemical composition of the film: 
PbBi1212 deposited on a SrTiO$_{3}$ (STO) substrate at low temperature forms an amorphous film which, at high temperature, 
crystallizes with the crystallographic symmetry of the substrate. 
We demonstrated that this technique works well for fabricating single-phase PbBi1212 epitaxial films. 
It is found that the PbBi1212 system exhibits superconductivity at approximately 50~K. 
We also clarify various superconducting properties of the Pb1212 system such as substitution effects, coherence lengths, and anisotropy. 
The two-step growth technique may provide a method to fabricate next-generation superconducting wires from ubiquitous elements.

\section{Crystal growth and experimental methods}
PbBi1212 epitaxial film was grown by a two-step technique consisting of a low temperature sputtering step and a high temperature $ex$ $situ$ growth step. 
Sputtering targets were synthesized by the solid-state reaction method using high purity powders ($>$~99.9\%) of 
PbO, Bi$_{2}$O$_{3}$, SrCO$_{3}$, Y$_{2}$O$_{3}$, CaCO$_{3}$, and CuO. 
These powders were mixed into compositions of (Pb$_{0.75}$Bi$_{0.25}$)$_{1.5}$Sr$_{2}$Y$_{1-x}$Ca$_{x}$Cu$_{2.6}$O$_{z}$ ($x=0.4-0.7$) 
and calcined two times: first at 860~$^\circ$C for 10~h in air and then at 880~$^\circ$C for 10~h in air. 
After calcination, the powders were pressed into cylindrical pellets 100~mm in diameter and 7~mm in height, 
and they were sintered at $1020-200x$~$^\circ$C for 24~h in air, where $x$ is the Ca concentration. 
For depositing the PbBi1212 amorphous films on STO (100) substrates, we used the following sputtering conditions: 
the sputtering gas pressure was 100~mTorr (60~sccm~Ar and 15~sccm~O$_{2}$), the anode voltage was 1.4~kV, 
and the substrate temperature was approximately 200~$^\circ$C (the substrates were not heated). 
The deposition time was set to 1 -- 2~h. 
The thickness of the PbBi1212 thin films measured by a stylus-based profilometer was 1800 -- 3500~\AA. 

PbBi1212 containers used for $ex$ $situ$ growth were made of polycrystalline pellets prepared in the same way as the sputtering targets at compositions of 
Pb$_{1-y}$Bi$_{y}$Sr$_{2}$Y$_{0.3}$Ca$_{0.7}$Cu$_{2}$O$_{z}$ ($y$~=~0~--~0.5). 
The mixed powders were calcined three times at 880~$^\circ$C for 10~h in air, 
pressed into two cylindrical pellets 26~mm in diameter and 5~mm in height, 
and sintered at 1007~$^\circ$C for 3~h in air. 
For epitaxial growth, amorphous films on STO substrates were placed in a pit (8~$\times$~8~$\times$~2~mm$^{3}$) formed 
at the center of one of the sintered pellets as shown in Fig.~1(a). 
The other pellet was used as a lid for the growth container. 
The container containing the amorphous film was heated in a muffle furnace at 970~$^\circ$C for 6~h under an O$_{2}$ atmosphere
and cooled to room temperature at a rate of 200~$^\circ$C/h. 
The concentration of the films was determined by energy dispersive x-ray spectroscopy (EDS), and it was found that the concentration of Pb and Bi drastically changes during the film growth. 
The concentration of Pb and Bi is dominated by the composition of the growth container, whereas
the concentrations of the other elements in the films are dominated by the composition of the sputtering target. 

As-grown films are not superconducting and a subsequent quenching treatment is necessary to make them superconducting. 
In this treatment, the film was placed in a quartz tube and heated at 815~$^\circ$C in air for 1~h. 
Within two seconds after removing the quartz tube from the furnace, it was placed in liquid nitrogen. 
This procedure increases the hole concentration and makes the film superconducting, as for the case of bulk polycrystalline PbCu1212, as reported by Maeda $et$ $al$\,\cite{Maeda}. 

The Hall coefficient and temperature dependence of the resistivity under a magnetic field were measured by the ac four-probe method with a physical properties measurement system (Quantum Design Co. Ltd.). 
The Hall coefficient was determined at various temperatures by a linear fit to the transverse resistivity as a function of the external magnetic field ($B$~$\parallel$~$c$) between $\pm$~5~T.

\section{Results and discussion}

\begin{figure}
\begin{center}
\includegraphics[width=85mm]{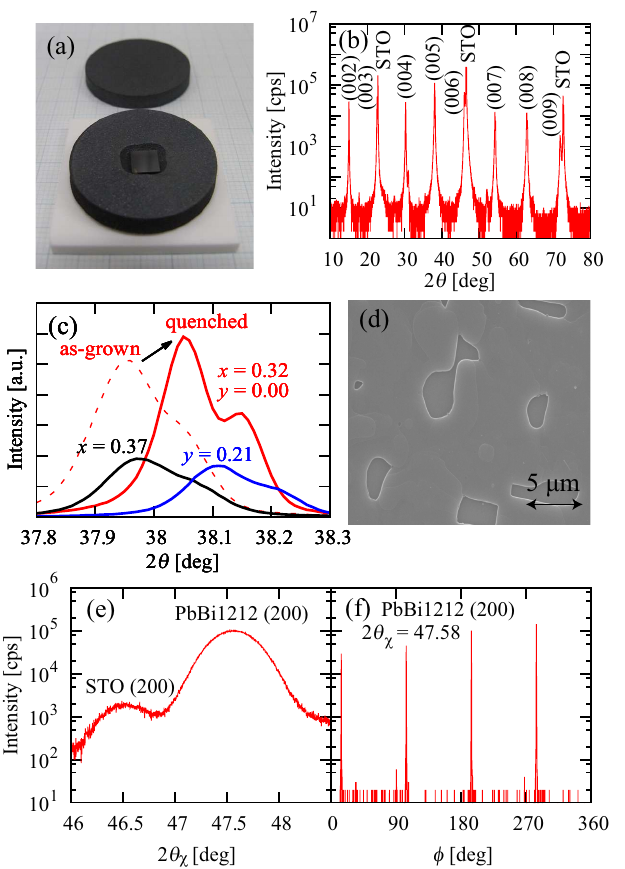}
\end{center}
\caption{
(a)~Photograph of the growth container and the amorphous film. 
(b)~Out-of-plane XRD $\theta$--2$\theta$ scan for single-phase Pb1212 thin film ($x$~=~0.32, $y$~=0.00). 
(c)~Magnified $\theta$--2$\theta$ scan at PbBi1212 (005) with different Ca and Bi concentrations. 
Red, blue, and black curves represent data for ($x$,~$y$) =~(0.32,~0.00), (0.32,~0.21), and (0.37,~0.00), respectively. 
$c$-axis lattice constants are 11.832~\AA\,, 11.820~\AA\,, and 11.861~\AA\,, respectively. 
The dashed (red) line represents data for the as-grown film ($c$~=~11.861~\AA) with ($x$,~$y$) =~(0.32,~0.00). 
(d)~Surface morphology for single-phase PbBi1212 thin film ($x$~=~0.31, $y$~=0.06). 
(e)~In-plane XRD 2$\theta$$_{\chi}$--$\phi$ scan pattern around (200) peaks of PbBi1212 and STO. 
(f)~$\phi$ scan pattern at PbBi1212 (200) (2$\theta$$_{\chi}$ = 47.58$^\circ$).}
\label{f1}
\end{figure}

\subsection{Crystallographic characterization}
Figure~1(b) shows the out-of-plane x-ray diffraction (XRD) $\theta-2\theta$ scan results for a thin film of a single-phase PbBi1212. 
These data indicate the complete $c$-axis alignment of the film. 
In thin films of single-phase PbBi1212, the peaks from other phases were completely absent or less than 0.5\% of the peak magnitude for PbBi1212 (005). 
It was found that a Ca concentration of $x<0.36$ and a Bi concentration of $y<0.20$ are required to obtain single-phase PbBi1212 films. 
Impurity phases such as Bi2212 were detected in samples whose concentrations of Ca or Bi exceeded these limits. 
To prevent the growth of the impurity phase, the Ca concentration of the sputtering target has to be less than 0.5 and the Bi concentration of the growth container has to be less than 0.25. 

Figure~1(c) represents the magnified $\theta-2\theta$ scan profiles at the peaks of PbBi1212 (005).
It is found that the quenching treatment reduces the $c$-axis lattice constant (approximately 0.03~\AA) and FWHM of the peak. 
This indicates that the lattice strains along the $c$ axis are reduced by the quenching treatment. 
The O$_{2}$ annealing at 500~$^\circ$C for 12~h results in doping holes for quenched films. 
However, the annealing yields little effect for as-grown films before the quenching treatment. 
This suggests that a significant amount of oxygen deficiency is induced by the quenching treatment. 
Hall effect measurements for samples before and after quenching revealed that the quenching treatment decreases the Hall coefficient significantly. 
We consider that the decrease in oxygen content effects a reduction in the lattice strains and 
this is the reason why the hole concentration of the film increases by the quenching treatment. 
Samples with low Ca and Bi concentrations often exhibit high crystallinity and the peaks of K$\alpha_1$ and K$\alpha_2$ are separately identified as shown in Fig.~1(c). 
Increases in $x$ (Ca concentration) and $y$ (Bi concentration) lead to an increase and decrease in the $c$-axis lattice constant, respectively.
A significant substitution of Ca and Bi tends to increase the FWHM of the peak and decrease the peak intensity, which are attributed to the lattice deformation. 

Figure~1(d) is an image of the surface of a thin film of single-phase PbBi1212. 
The image was acquired by scanning electron microscopy. 
The surface morphology of the film is smooth except for a few steps. 
This morphology implies that film growth progresses along the surface of the substrate. 
Note that no cracks, grain boundaries, or impurity phase precipitations were found on the entire surface of the film. 

We used in-plane XRD measurements to check whether the films grow epitaxially on the substrates. 
Figure~1(e) shows the result of a 2$\theta$$_{\chi}$--$\phi$ scan for a single-phase PbBi1212 film. 
The peaks of STO (200) and PbBi1212 (200) are identified. 
This result shows that lattice relaxation occurs from the 2\% in-plane lattice mismatch between STO ($a$~=~3.905~\AA) and PbBi1212 ($a$~=~3.81--3.83~\AA). 
The result of an in-plane $\phi$ scan at the peak of PbBi1212 (200) (2$\theta$$_{\chi}$~=~47.58$^\circ$) is displayed in Fig.~1(f). 
Four peaks are found at 90$^\circ$ between each peak, and the difference in the angle of the peak between STO and PbBi1212 is less than 0.7$^\circ$. 
These results indicate that the PbBi1212 thin films grow epitaxially on STO (100) substrates. 

\begin{figure}[htbp]
\begin{center}
\includegraphics[width=85mm]{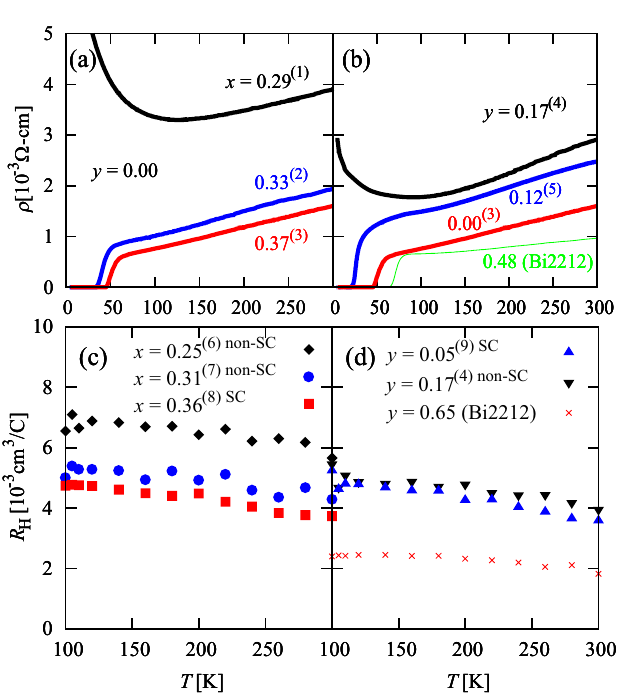}
\end{center}
\caption{Temperature dependence of resistivity and Hall coefficients for samples with different Ca and Bi concentrations. 
(a)~Black, blue, and red curves represent data for ($x$,~$y$) =~(0.29,~0.00), (0.33,~0.00), and (0.37,~0.00), respectively. 
(b)~Black, blue, red, and green curves represent data for ($x$,~$y$) =~(0.35,~0.17), (0.35,~0.12), (0.37,~0.00), and (0.49,~0.48), respectively. 
(c)~Black diamonds, filled blue circles, and red squares represent data for ($x$,~$y$) =~(0.25,~0.15), (0.31,~0.13), and (0.36,~0.17), respectively. 
(d)~Black triangles, blue triangles, and red crosses represent data for ($x$,~$y$) =~(0.35,~0.05), (0.35,~0.17), and (0.50,~0.65), respectively. 
Numbers in parentheses correspond to those in Fig.~3.} 
\end{figure}
\begin{figure}
\begin{center}
\includegraphics[width=85mm]{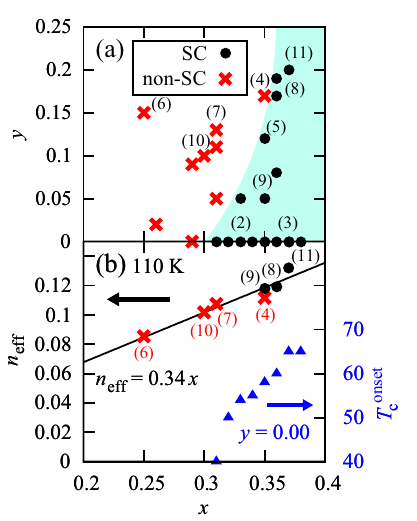}
\end{center}
\caption{(a)~Distribution of Ca and Bi concentrations in PbBi1212 epitaxial films and their superconductivity. 
Filled black circles correspond to superconducting samples and (red) crosses correspond to nonsuperconducting samples. 
The hatched area shows the range of concentrations for which superconducting films are obtained. 
(b)~Effective hole density $n_{\rm eff}$ [(red) crosses and filled black circles, left axis] for various Bi concentrations $y$ 
and $T_{\rm c}$$^{\rm onset}$ [(blue) triangles, right axis] for $y=0$ as a function of Ca concentration $x$. }
\label{f1}
\end{figure}

\subsection{Substitution effect}

Figures~2(a) and 2(b) show the temperature dependence of the resistivity for samples with different Ca and Bi concentrations, respectively. 
The composition of the film was determined by EDS at three points (25~$\times$~25~$\mu$m$^2$ for each point). 
As shown in Fig.~2(a), an increase in Ca concentration $x$ leads to a decrease in resistivity and $x>0.3$ is necessary for the superconducting transition. 
Notably, a very small increase in Ca concentration induces a transition from an insulator to a superconductor. 
However, further increasing the Ca concentration does not significantly affect $T_{\rm c}$. 
In contrast, replacing Pb with Bi up to $y$~=~0.17 drastically decreases $T_{\rm c}$, as shown in Fig.~2(b). 
For $y$~$>$~0.20, the films include Bi2212 as a secondary phase and exhibit distinctly different properties than the single-phase films. 
The resistivity decreases with increasing Bi concentration and even without the quenching treatment, a sharp superconducting transition takes place at 80~K. 
Also, the XRD peaks of the Bi2212 phase become more intense. 
Therefore, we conclude that the superconducting transition observed for $y$~$>$~0.20 is due to the Bi2212 impurity phase. 

The temperature dependence of the Hall coefficient for different Ca and Bi concentrations is shown in Figs.~2(c) and 2(d). 
Figure~2(c) suggests that replacement of Y with Ca increases the hole concentration. 
However, as shown in Fig.~2(d), no significant change in the Hall coefficient as a function of the Bi concentration is observed for the single-phase samples ($y<0.20$). 
The significant decrease in the Hall coefficient for $y>0.20$ is attributed to the Bi2212 impurity phase. 
We find that the slight increase in hole concentration, e.g., from $x$~=~0.31 to 0.36 shown in Fig.~2(c), causes superconductivity. 
All nonsuperconducting samples are insulators at low temperatures. 
This sudden change from insulator to superconductor is quite particular to this PbBi1212 system. 
In contrast, Bi substitution does not introduce holes, but it introduces lattice deformation that strongly decreases $T_{\rm c}$, especially at $x$~$\approx$~0.32. 
This is consistent with the $\rho(T)$ data in Fig.~2(b), 
where the residual resistivity [determined by the intersection of the linear extrapolation of $\rho(T)$ data above $T_{\rm c}$ and $T$ = 0 axis] increases with $y$. 
The result of the XRD measurement in Fig.~1(c) also shows the influence of lattice deformation from the Bi substitution. 
Furthermore, the reduction in the $c$-axis lattice constant due to the Bi substitution presumably causes the decrease in $T_{\rm c}$, 
because the $T_{\rm c}$ of the 1212 family tends to decrease with decreasing distance between adjacent CuO$_2$ double layers ($T_{\rm c,~Hg1212}$ $>$ $T_{\rm c,~Tl1212}$ $>$ $T_{\rm c,~Pb1212}$) as seen in other families. 

The substitution effect found in this study is summarized in Fig.~3(a). 
This plot clearly illustrates that, for superconductivity in PbBi1212, a Ca concentration $x$~$>$~0.3 is essential and a lower Bi concentration is preferable. 
The effective number of holes per Cu atom ($n_{\rm eff}=V_{\rm Cu}/R_{\rm H}e$) as a function of the Ca concentration $x$ for various Bi concentrations $y$ is plotted in Fig.~3(b), 
where $V_{\rm Cu}$ is the unit cell volume for a Cu atom. 
The quantity $n_{\rm eff}$ depends linearly on $x$, irrespective of $y$. 
Among these samples ($y>0.05$), the superconductivity occurs at $n_{\rm eff}=0.12$, which is between the hole number for optimum YBa$_2$Cu$_3$O$_{7-\delta}$ (YBCO) ($T_{\rm c}=90K$, $n_{\rm eff}=0.28$) 
and the hole number for the superconductor-insulator transition ($n_{\rm eff}=0.027$)\,\cite{Semba}. 
This trend is the same as for La$_{2-x}$Sr$_x$CuO$_4$ (LSCO)\,\cite{Ando}. 
Therefore, we expect a further increase in $T_{\rm c}$ for our sample upon doping with more hole carriers. 
In fact, we find that, for samples with $y$~=~0, $T_{\rm c}$$^{\rm onset}$ increases with $x$ up to $x$~=~0.37, as shown in Fig.~3(b). 
The linear fit shown in Fig.~3(b) suggests that carrier doping is governed by the substitution of Y$^{3+}$ for Ca$^{2+}$ and 0.34 hole is introduced by the substitution of a Ca$^{2+}$ ion.  
This is similar to the underdoped region in LSCO\,\cite{Takagi} and YBCO\,\cite{Semba}, where the effective number of holes is well described as $n_{\rm eff}$~$\propto$~$x$. 
The remarkable point of doping in Pb1212 is that the fitting line is extrapolated to the zero point of $n_{\rm eff}$ for $x$ $\to$ 0. 
This indicates that all the transport carriers originate from the Ca substitution, and all the doped carriers contribute as the transport carriers. 

As $x$ is increased above 0.37, $T_{\rm c}$ remains constant, whereas at 300~K, $\rho$ decreases. 
This implies that increasing $x$ creates not only hole carriers but also lattice deformation that is more sensitive to the emergence of the superconductivity than to the normal-state conductivity. 
The broadened (005) peak shown in Fig.~1(c) elucidates that the Ca substitution introduces crystallographic disorder as discussed above. 
The (Y,~Ca) site of Pb1212 is inside the CuO$_2$ double layer, as is the case for (Y,~Ca)Ba$_2$Cu$_3$O$_7$ with $T_{\rm c}$ approximately equal to 90~K. 
Eisaki~$et$~$al$. pointed out the strong influence of the (Y,~Ca) site disorder on $T_{\rm c}$\,\cite{Eisaki}. 
Concerning Ca substitution at $y=0$, we conclude that chemical disorder prevents the increase in $T_{\rm c}$ expected from hole doping. 
To obtain the highest $T_{\rm c}$ for this material, the film properties must be further optimized.

\begin{figure}[htbp]
\begin{center}
\includegraphics[width=85mm]{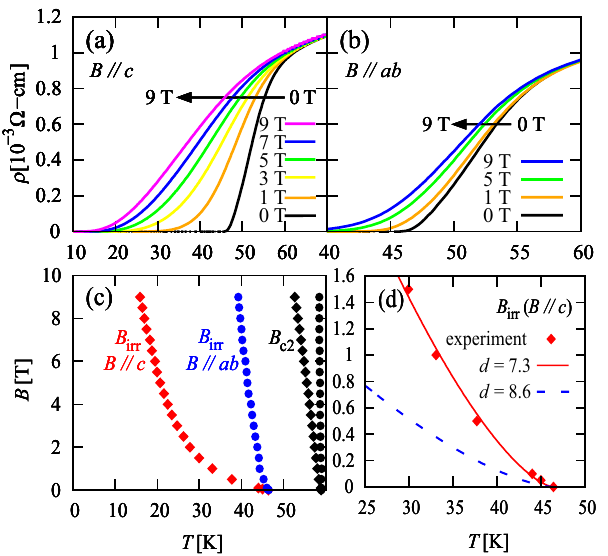}
\end{center}
\caption{Temperature dependence of resistivity for (a)~$B$~$\parallel$~$c$ and (b)~$B$~$\parallel$~$ab$. 
(c)~Irreversibility fields and upper critical fields in a sample with ($x$,~$y$) =~(0.34,~0.00). 
(c)~Diamonds and filled circles represent data for $B$~$\parallel$~$c$ and $B$~$\parallel$~$ab$, respectively. 
(d)~Solid and dashed lines are obtained from the empirical formula with $d$~=~7.3 and 8.6, respectively.
}
\end{figure}

\subsection{Superconducting anisotropy}
Figures~4(a) and 4(b) show the temperature dependence of resistivity in a magnetic field of up to 9~T for $B$~$\parallel$~$c$ and $B$~$\parallel$~$ab$ for the Bi-free ($y=0.00$) sample. 
The zero-field transition is significantly broadened by applying $B$~$\parallel$~$c$, whereas the transition is broadened only slightly for $B$~$\parallel$~$ab$. 
No sharp transition, indicating the vortex lattice melting transition\,\cite{Kwok}, is observed. 
This result is attributed to the distribution of the superconducting condensation energy accompanied by an 
inhomogeneous oxygen concentration and lattice disorder in the film. 

To discuss the superconducting anisotropy, the coherence lengths along the $ab$ plane $\xi_{ab}(0)$ and the $c$ axis $\xi_{c}(0)$ are estimated from the upper critical field $B_{\rm c2}$, 
which is defined by the criterion of $\rho$/$\rho_{\rm n}$~=~90\%. 
This criterion excludes the influence of the flux flow on $B_{\rm c2}$.
$B_{\rm c2}$ for $B$~$\parallel$~$c$ and $B$~$\parallel$~$ab$ is plotted in Fig.~4(c) as a function of temperature. 
The quantities $B_{\rm c2}^{c}(0)$ and $B_{\rm c2}^{ab}(0)$ are estimated to be 54 and 492~T, respectively, using the Werthamer-Herfand-Hohenberg theory\,\cite{Werthamer}, 
which gives $B_{\rm c2}(0)=-0.69T_{\rm c}(dB_{\rm c2}/dT)|_{T_{\rm c}}$. 
According to the relation $B_{\rm c2}=\Phi_{0}/2\pi\xi^{2}$, the coherence lengths $\xi$ were derived as $\xi_{ab}(0)$~=~25~\AA\, and $\xi_{c}(0)$~=~2.7~\AA. 
For $B$~$\parallel$~$c$, the tangent $(dB_{\rm c2}/dT)|_{T_{\rm c}}=-1.32$~T/K is slightly smaller than that of the YBCO single-crystal (--1.9~T/K)\,\cite{Welp}. 
This result indicates that the mean free path $l$ of Pb1212 is slightly longer than that of YBCO, because $dB_{\rm c2}/dT$ is roughly proportional to $l^{-1}$ in the dirty limit. 
This suggests that Pb1212 is slightly cleaner than YBCO ($\xi_{ab}(0)$=16~\AA). 
Meanwhile, the value of $\xi_c(0)$ is close to the thickness of superconducting layers of Pb1212 (3.2~\AA). 
This is a common feature among cuprates except for the Bi-family. 

Consequently, the anisotropy, $\gamma=\xi_{ab}(0)/\xi_{c}(0)$ is estimated to be 9.2. 
This anisotropy is similar to that of bulk Tl1212 ($\gamma$~$\sim$~10)\,\cite{Warmont} and Hg1212 ($\gamma$~$\sim$~7.7)\,\cite{Kim}. 
A further reduction in anisotropy was observed for Bi-doped samples (e.g.,~$\gamma$~$=$~6.5 with $y$~$=$~0.12), 
although the $T_{\rm c}$ values of these samples are lower than those of Bi-free samples as discussed above. 
The reduced anisotropy is presumably caused by the decrease in the distance between adjacent CuO$_2$ double layers 
because the ionic radius of Bi$^{5+}$ is smaller than that of Pb$^{4+}$\,\cite{Ehmann}. 
These results reveal that the anisotropy of PbBi1212 is much less than that of the modulation-free PbBi2212 ($\gamma$~$\sim$~25)\,\cite{Gladyshevskii}. 
This implies that replacing the double (Pb,~Bi)O block layers with monolayers results in a reduction in the superconducting anisotropy and an enhancement in the coupling between CuO$_2$ layers.

In Fig.~4(c), the irreversibility field $B_{\rm irr}$ defined by the criterion $\rho$~$=$~$1\times10^{-5}$~$\Omega$-cm is also plotted. 
Unlike the case of $B_{\rm c2}$, $B_{\rm irr}$ (especially for $B$~$\parallel$~$c$) is significantly shifted to the low temperature side in a high magnetic field. 
Figure~4(d) shows an enlarged plot of $B_{\rm irr}$ ($B$~$\parallel$~$c$) for $B$~$<$~1.6~T. 
It also shows the results of fitting the data to the empirical function \\
$B_{\rm irr} [\rm T]=2\times 10^{3}{\rm exp}(-0.78$$d)(1-T/T_{\rm c})^{1.5}$ ($T>0.7T_{\rm c}$)\,\cite{Shimoyama}, 
where $d$~(\AA\,)\,is the distance between adjacent CuO$_2$ planes. 
The curve fit gives the value $d=7.3$~\AA\,, which is less than the result of $d=8.6$~\AA\, from crystal structure refinements 
based on XRD and neutron diffraction\,\cite{Glaser}. 
Based on EDS analysis of the film, a considerable portion of Pb ions in the block layer is possibly substituted by Cu ions. 
Assuming that the substituted sites, which form CuO$_6$ octahedral sites, accumulate and form CuO$_2$ single layers and thereby partly spread the film, 
the effective value of $d$ may be considerably reduced. 
In YBCO, the empirical formula applies for $d$~=~4.15~\AA\, which is close to the distance between the CuO$_2$ plane and CuO chain\,\cite{Shimoyama}. 

\section{Summary}
The superconductivity of PbBi1212 was vertified by measuring the anisotropic transport properties in epitaxial films prepared by the two-step growth technique. 
Films of single-phase PbBi1212 were obtained with a Ca concentration of less than 0.36 and a Bi concentration of less than 0.20. 
We found that a Ca concentration greater than 0.3 and a quenching treatment are essential for making the epitaxial films superconducting. 
Increasing the Ca concentration $x$ leads to an increase in the hole density ($n_{\rm eff}$ = $0.34\,x$), and increasing the Bi concentration leads to a decrease in the critical temperature $T_{\rm c}$. 
A significant substitution of Ca and Bi induces the lattice deformation, which strongly decreases $T_{\rm c}$. 
The highest $T_{\rm c}$$^{\rm onset}$ found for single-phase PbBi1212 films is 65~K, 
and for the multiphase film, the drop in resistivity starting at a higher temperature is attributed to the Bi2212 impurity phase. 
The coherence lengths of Pb1212 estimated from the upper critical field are 25 and 2.7~\AA\, along the $ab$ plane and $c$ axis, respectively. 

\begin{acknowledgements}
This work was partly supported by the Global COE project ``Education and Research on Photonics and Electronics Science and Engineering" at Kyoto University. 
\end{acknowledgements}

\end{document}